\documentclass[11pt,preprint]{aastex}
\usepackage{chngpage}
\usepackage{graphicx}
\usepackage{natbib}
\shorttitle{Possible explanation for PSR J0738-4042}
\shortauthors{Yu \& Huang}

\begin{document}

\title{Reduced spin-down rate of PSR J0738-4042 explained as due to an asteroid disruption event}

\author{Y. B. Yu\altaffilmark{1, 2} \& Y. F. Huang\altaffilmark{1, 2}}

\altaffiltext{1}{Department of Astronomy, Nanjing University, Nanjing 210046, China; hyf@nju.edu.cn}
\altaffiltext{2}{Key Laboratory of Modern Astronomy and Astrophysics (Nanjing University), Ministry of Education, Nanjing 210046, China}

\begin{abstract}
Long term observations by Brook et al. reveal that the derivative of rotational frequency of PSR J0738-4042 changed abruptly in 2005.
Originally, the spin-down rate was relatively stable, with the rotational frequency derivative of $-1.14 \times 10^{-14}~\rm s^{-2}$. After September 2005, the
derivative began to rise up. About 1000 days later, it arrived at another relatively stable value of about $-0.98 \times 10^{-14}~\rm s^{-2}$, indicating that the pulsar is spinning-down relatively slowly.
To explain the observed spin-down rate change, we resort to an asteroid disrupted by PSR J0738-4042. In our model, the orbital angular
momentum of the asteroid is assumed to be parallel to that of the rotating pulsar, so that the pronounced reduction in the spin-down
rate can be naturally explained as due to the transfer of the angular momentum from the disrupted material to the central pulsar. The derived
magnetospheric radius is about $4.0 \times 10^{9}$ cm, which is smaller than the tidal disruption radius ($4.9 \times 10^{10}$ cm).
Our model is self-consistent. It is shown that the variability of the spin-down rate of PSR J0738-4042 can be quantitatively accounted for by the
accretion from the asteroid disrupted by the central pulsar.
\end{abstract}


\section{Introduction}
\label{sect:intro}

Pulsars are widely believed to be fast rotating neutron stars, which are compact objects with typical radius ${R \sim 10^{6}}$ cm and mass ${M \sim 1.4M_{\odot}}$.
Pulsars usually have relatively stable pulse profiles and rotational periods, due to which they can even be used as unique high-precision clocks in
experimental astrophysics. However, sometimes there are some timing variabilities observed in pulsars, such as glitches (Wang et al. 2000; Yuan
et al. 2010; Espinoza et al. 2011; Manchester \& Hobbs 2011; Yu et al. 2013), micro-glitches (Cognard \& Backer 2004; Mandal et al. 2009), pulse
profile changing (Rankin 1986; Burgay et al. 2005; Poutanen et al. 2009; Karastergiou et al. 2011; Bilous et al. 2014), pulse nulling (Deich et
al. 1986; Rankin \& Wright 2008; Jones 2011; Li et al. 2012), and pulse drifting (Backer 1973; Page 1973; Ruderman 1976; Esamdin et al. 2005;
Jones 2014).

Especially, neutron star glitches, characterized by a sudden increase and gradual relaxation of rotation frequency,
are very interesting astrophysical phenomena and have been observed from many normal pulsars and magnetars (Kaspi et al. 2000; Dib et al. 2009;
Livingstone et al. 2010; Gavriil et al. 2011), where magnetars are a type of pulsars with dipole magnetic fields significantly stronger than
$4.4 \times 10^{13}$ G (Usov 1992; Ducan \& Thompson 1992; Olausen \& Kaspi 2014). Usually, the glitch is explained to be associated with sudden
decoupling of the pinned vortex lines in the crustal neutron superfluid region (Anderson \& Itoh 1975; Pines \& Alpar 1985; Alpar \& Baykal 2006;
Warszawski \& Melatos 2011; Haskell et al. 2012; Warszawski et al. 2012; Chamel 2013). In the normal steady state, the pinned superfluid is coupled to the rest
of the neutron star. But when the pinned superfluid is decoupled due to continuously increasing rotation lag, the angular momentum of the fast-rotating
interior superfluid component will be transferred to the outer solid crust, leading to an observed glitch. There are also some other models to explain the
observed glitches, such as the platelet collapse model (Morley \& Schmidt 1996), the superfluid r-mode instability mechanism (Glampedakis \& Andersson 2009),
the snowplow model (Seveso et al. 2012), and the starquake model (Zhou et al. 2014).

Anti-glitches were also observed in pulsars, such as from 1E 2259+586 (Archibald et al. 2013) and 1E 1841-045 (Sasmaz Mus et al. 2014). Anti-glitches could be
generated by either an internal mechanism such as an impulsive angular momentum transfer between the superfluid region and the crust (Thompson et al. 2000),
or an external mechanism such as a sudden twisting of the magnetic field lines (Lyutikov 2013) or accretion of retrograde matter (Katz 2014; Ouyed et al.
2014). However, the anti-glitch of the magnetar 1E 2259+586 was very special, associated with a hard X-ray burst, which strongly challenges traditional
glitch theories. Huang \& Geng (2014) proposed a completely different model to interpret this strange behavior. In their model, the sudden spin-down is
explained to be induced by collision of a small solid body with the central magnetar. The associated hard X-ray burst and the decaying softer
X-ray emission can be explained well.

Recently, a sudden change in the spin-down rate of PSR J0738-4042 was reported by Brook et al. (2014). The value of the spin-down rate was originally quite stable.
But after September 2005, a significant reduction in the spin-down rate was observed. This abrupt change was coincided with the appearance of a new component in the
average pulse profile (Karastergiou et al. 2011). As these timing and emission properties can not be explained by normal intrinsic pulsar processes, Brook et al. (2014)
argued that they were generated by an external mechanism and invoke an asteroid disrupted by the central neutron star. However, we noticed that their
calculations are not self-consistent. Here we show that accretion from tidally disrupted material is truly a possible explanation for the special behaviors observed in
the spin-down rate and present a self-consistent calculation.

The structure of our paper is organized as follows. We summarize the observational facts of PSR J0738-4042 in Section 2. The asteroid disruption model,
including the tidal disruption and accretion processes, are described in Section 3. We summarize our results in the final section with a brief discussion.

\begin{figure}
   \begin{center}
   \plotone{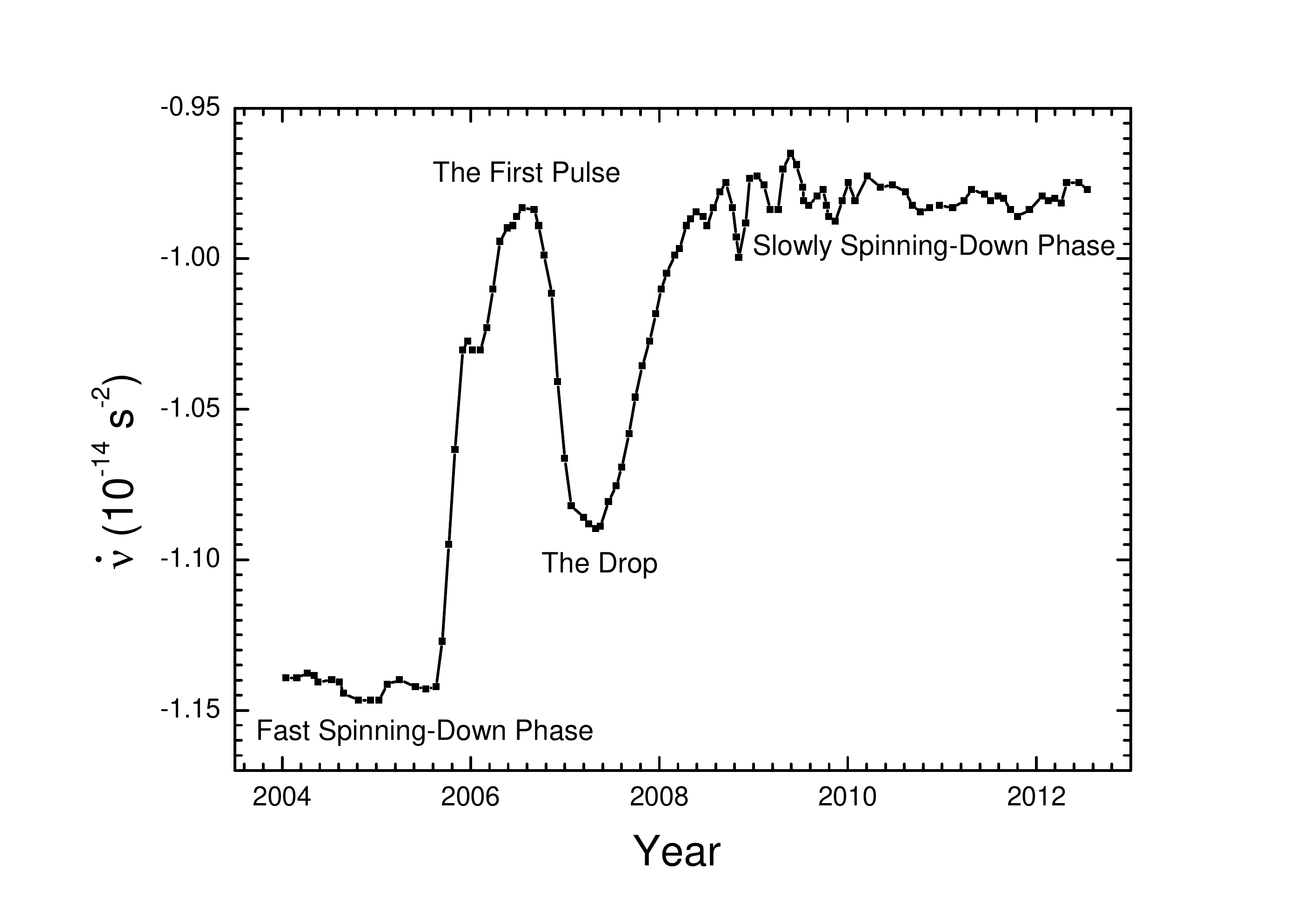}
   \caption{A schematic illustration of the evolution of the rotational frequency derivative ($\dot{\nu})$
   of PSR J0738-4042. The observational data are taken from Brook et al. (2014). See Brook et al. (2014)
   for a detailed plot.}
   \label{Fig:plot1}
   \end{center}
\end{figure}

\section{Observational facts}
\label{sect:obs}

PSR J0738-4042 is a radio-emitting neutron star with a rotational period of 0.375 s and spin-down evolution of $\dot{\nu} = -1.14 \times 10^{-14}$
${\rm s^{-2}}$ (Brook et al. 2014), where $\dot{\nu}$ is the time derivative of the rotational frequency $\nu$, which is gradually decreasing due to
magnetic dipole radiation. As the target of a long-term monitoring campaign, PSR J0738-4042 has been observed by researchers at the Hartebeesthoek Radio
Astronomy Observatory in South Africa from September 1988. PSR J0738-4042 has also been observed by the Parkes radio telescope in Australia as part of
the Fermi timing programme since 2007 (Weltevrede et al. 2010). PSR J0738-4042 shares similar rotational properties with other isolated, middle-aged radio
pulsars. The pulse profile and timing property of PSR J0738-4042 are originally stable, with
$\dot{\nu} = -1.14 \times 10^{-14}$ ${\rm s^{-2}}$ (see Fig. 1). In Fig. 1, since a smaller $\dot{\nu}$ value means
the pulsar is spinning-down relatively quickly, we call this initial stage as the Fast Spinning-Down Phase. However, from September 2005, a dramatic change in the spin-down rate occurred (Brook et al. 2014, also see Fig. 1 for a
schematic illustration). This change was accompanied by a new radio component that drifted
on the leading edge of the pulse profile, which was first noted by Karastergiou et al. (2011). The value of $\dot{\nu}$ gradually rose up to a peak value of
$-0.98 \times 10^{-14}~\rm s^{-2}$ about 450 days later since September 2005. The amplitude of this change in the spin-down rate is about $0.16 \times 10^{-14}~\rm s^{-2}$.
Note that there are also some small variabilities during the rising phase. As marked in Fig. 1, we call this stage as The First Pulse. After reaching the peak value, $\dot{\nu}$ began to drop down slightly. In Fig. 1, this stage is marked by ``The Drop".
After dropping down to a value of $-1.08 \times 10^{-14}~\rm s^{-2}$, another significant change in the spin-down rate began. The value of $\dot{\nu}$ rose back to
$-1.0 \times 10^{-14}~\rm s^{-2}$ in about 300 days. After some small turbulence, the spin-down rate finally arrived at another relatively stable phase, with a new
value of $\dot{\nu} = -0.98 \times 10^{-14}~\rm s^{-2}$ (Brook et al. 2014). We call this stage the Slowly Spinning-Down Phase, since a higher $\dot{\nu}$ value indicates that the pulsar is spinning-down relatively slowly.

As the change in the spin-down rate of PSR J0738-4042 is accompanied by an emergent radio component that drifts with respect to the rest of the pulse profile, normal
intrinsic pulsar processes cannot explain these radio emission and timing features. Brook et al (2014) suggested that they are witnessing an encounter of
the pulsar with an asteroid. Due to orbital perturbations and collisions, asteroids may migrate inwards and interact with the magnetosphere after being disrupted by
the central pulsar, affecting the pulse profile and the rotational stability. According to Brook et al. (2014), the change in the spin-down rate was caused by mass
supplied to the pulsar. However, when calculating the total accreted mass, they firstly related the reduction in the outflowing charge density to the change in the
spin-down rate. Then they multiplied the reduced charge density with the speed of light, the polar cap area, and the duration of the new spin-down state to get a
mass of about $1.0 \times 10^{15}$ g. They pointed out that this value is the mass of the asteroid encountering the central pulsar. But in fact, the mass estimated
by Brook et al. (2014) is the reduction mass in the total outflowing plasma above the polar caps. It cannot be regarded as the mass of the accreted material, which
itself is a kind of inflow. Thus it is obvious that their interpretation and calculations are not self-consistent. In this paper, to explain the behavior of the
spin-down rate of PSR J0738-4042, we reconsider the tidal disruption and accretion processes in detail and give a self-consistent modeling.

\section{Model}
\label{sect:model}

Interestingly, Shannon et al (2013) showed that the timing variabilities of PSR B1937+21 is consistent with the existence of an asteroid belt. Actually, the planetary
and disk systems have been confirmed in some neutron stars, such as PSR B1257+12 (Wolszczan 1994), PSR B1620-26 (Thorsett et al. 1999) and
the magnetar 4U 0142+61 (Wang et al. 2006). The mechanism of collision between small bodies and neutron stars has also been widely used to interpret
transient X/$\gamma$-ray events (van Buren 1981; Livio \& Taam 1987; Zhang et al. 2000; Campana et al. 2011). To explain the pronounced change in the spin-down rate of
PSR J0738-4042, we invoke an close encounter between an asteroid and the central pulsar. Since the spin-down rate reduced significantly (the $\dot{\nu}$ value increased, see Fig. 1) after September 2005,
the orbital angular momentum of the asteroid is assumed to be parallel to that of PSR J0738-4042. Firstly, the asteroid will be disrupted
at the tidal disruption radius. After the disruption, some fraction of the disrupted material will be ejected at high speed, while the rest is bound to
the central pulsar (Rees 1988). When the bound material moves around the central pulsar, the gaseous debris will be partially accreted on to the surface of the
central pulsar, changing the emission and pulse profile signatures. Initially, the bound orbit is highly eccentric. As the material tends to stay in an orbit
with the lowest energy for a given angular momentum, i.e. a circular orbit, the disrupted and bound debris will experience a process called the orbital circularization.
After the orbital circularization, a disk would be formed. From then on, the mass accretion rate tends to be somewhat constant before the exhaustion of the material in
the transient disk. We show that the evolution behavior of the spin-down rate of PSR J0738-4042 can be explained by the accretion processes at different stages.

\subsection{Tidal disruption and initial accretion}

Due to strong gravitational force, the asteroid that comes too close to the pulsar will be broken up when coming to the tidal disruption radius given by (Hills 1975)
\begin{equation}
R_{\rm t} \approx (6M/\pi\rho)^{1/3},
\end{equation}
where $M$ is the pulsar mass and $\rho$ is the characteristic central density of the asteroid, which will be taken as $8\rm ~g~cm^{-3}$ for a typical homogeneous iron-nickel body
(Colgate \& Petschek 1981) for simplicity in our calculations. Thus the tidal disruption radius can be calculated to be $R_{\rm t} = 4.9 \times 10^{9} \rm cm$ typically.
After the disruption, part of the disrupted material will be bound to the central pulsar. In the first flyby, a small portion of the bound material will be accreted
towards the central pulsar. During the accretion process, the infalling material will exert its own force to generate the ram pressure. Because of the existence of
the dipole magnetic field, there will also be a magnetic pressure at any given radius. Equating the two pressures, we can get the $\rm Alfv\acute{e}n$ radius, which is also known as the magnetospheric radius,
given by
\begin{equation}
r_{\rm m} = \left(\frac{\mu^{4}}{GM{\dot{m}}^{2}}\right)^{1/7},
\end{equation}
where G is the gravitational constant and $\dot{m}$ is the mass accretion rate. $\mu = B_{0}R^{3}$ is the
magnetic dipole moment of the pulsar and $B_{0}$ is the surface magnetic field, which can be estimated as
$B_{0} = 3.2 \times 10^{19} (-\dot{\nu}/\nu^{3})^{1/2} \approx 8 \times 10^{11}$ G for PSR J0738-4042, with $\nu$ and $\dot{\nu}$ being 2.667 $\rm s^{-1}$ and
$-1.15 \times 10^{-14}~\rm s^{-2}$ respectively (see the Fast Spinning-Down Phase in Fig. 1). In our calculations, the disrupted material is assumed to be accreted and adhered to the central pulsar from the
magnetospheric radius. Conservation of angular momentum then leads to
\begin{equation}
I\cdot2\pi\nu + mVr_{\rm m} = I\cdot2\pi(\nu + \Delta\nu),
\end{equation}
where $I$ is the moment of inertia of the central pulsar, which is taken as a typical value of $I \sim 10^{45} \rm g~cm^{2}$ (Pizzochero 2011;
Hooker et al. 2013), and $m$ is the mass of the accreted material. $V$ is the velocity of the asteroid at the radius $r_{\rm m}$, assumed to be $V = (2GM/r_{\rm m})^{1/2}$.
Through some simple derivation we can further simplify Equation (3) to
\begin{equation}
\dot{m}(2GMr_{\rm m})^{1/2} = 2\pi I\cdot\Delta(\dot{\nu}).
\end{equation}
To estimate the characteristic mass accretion rate, we firstly assume it to be constant. $\Delta(\dot{\nu})$ is taken as $0.16 \times 10^{-14}~\rm s^{-2}$,
which is determined from observational data (i.e., the difference of $\dot{\nu}$ in the Fast Spinning-Down Phase and in The First Pulse phase, see Fig. 1). For typical parameters of $M = 1.4 M_{\odot}$ and $R = 10^{6}$ cm, the average mass accretion rate can be calculated
as $\dot{m} = 1.4 \times 10^{13}~\rm g~s^{-1}$. In this case, the $\rm Alfv\acute{e}n$ radius can be estimated as $r_{\rm m} = 4.0 \times 10^{9} \rm cm$.

We argue that The First Pulse phase beginning from September 2005 in Fig. 1 is due to the initial accretion of the gaseous debris in the first flyby
after the disruption. Based on observations, there is a drop in the $\dot{\nu} - t$ diagram after The First Pulse, as shown in Fig. 1. This is because, after passing the pericenter,
the majority of the disrupted material will fly away from the central pulsar along the highly elliptical orbit. It will result in a decrease
of the mass accretion rate, leading to a drop of $\dot{\nu}$. In our calculations, we assume the duration of The First Pulse, which is about 700 days,
as the initial orbital period ($T$) of the bound debris. According to the formula $T = 2\pi\sqrt{\frac{a^{3}}{GM}}$, the semimajor axis $a$ of the initial elliptical
orbit can be calculated as $a = 2.3 \times 10^{13}$ cm, which is larger than the tidal disruption radius and the $\rm Alfv\acute{e}n$ radius, indicating that
our calculations are self-consistent. Assuming that the initial orbital period is also roughly the time for the bound debris to return to the pericenter,
we further have (Ulmer 1999; Lu et al. 2006)
\begin{equation}
T = \frac{2\pi R_{\rm p}^{3}}{(GM)^{1/2}(2r)^{3/2}},
\end{equation}
where $r$ is the radius of the asteroid and $R_{\rm p}$ is the pericenter distance. From this equation, the pericenter distance can be estimated as $R_{\rm p} = 1.48 \times 10^{10}$ cm.

\subsection{Circularization and stable accretion}

When the disrupted material moves around the central pulsar in elliptical orbit with high ellipticity, there will be
dissipative processes, e.g. collisions, shocks, viscous dissipation, etc (Shakura \& Sunyaev 1973). These processes will convert some
of the energy of the ordered bulk orbital motion into internal energy, part of which will be radiated and therefore lost
from the gas. Thus the gas has to sink deeper into the potential of the central pulsar, orbiting it more closely.
This in turn requires the gas to lose angular momentum. So most of the disrupted material will spiral towards the central pulsar
through a series of elliptical orbits with continuously decreasing ellipticities. The above process is called the orbital circularization.
During the circularization, some gas will move inwards due to the alpha viscosity process. In this study, as the corotation radius
$r_{\rm c} = (GM/\Omega^{2})^{1/3} \approx 2.2 \times 10^{8} \rm cm$ is smaller than the $\rm Alfv\acute{e}n$ radius
($4.0 \times 10^{9} \rm cm$), the gaseous debris will firstly accumulate at the $\rm Alfv\acute{e}n$ radius of the
pulsar. When the pressure exerted by the accreted material exceeds the magnetic pressure from the central pulsar, the gas
will stream on to the poles of the pulsar along the magnetic field lines, changing the spin and emission properties of the
central PSR J0738-4042. After the circularization, an accretion disk would be formed around PSR J0738-4042.
The observed relatively stable spin-down rate at the last stage in Fig. 1 (i.e., the Slowly Spinning-Down Phase) can be explained by the constant accretion from
the disk. Comparing $\dot{\nu}$ in the Fast Spinning-Down Phase with that in the Slowly Spinning-Down Phase, we again get the difference as $\Delta(\dot{\nu})= 0.16 \times 10^{-14}~\rm s^{-2}$. This again gives the accretion rate as $\dot{m} = 1.4 \times 10^{13}~\rm g~s^{-1}$. During the accretion timescale, i.e. from September 2005 to right now, the total accreted mass should be
$3.5 \times 10^{21}$ g, which is consistent with the mass range of asteroids around neutron stars.
In this study, we regard $3.5 \times 10^{21}~\rm g$ as the lower limit of the mass of the asteroid.

The total time for the bound debris to complete the circularization process around the central compact object is (Ulmer 1999; Lu et al. 2006)
\begin{equation}
t_{\rm cir} = n_{\rm orb}T,
\end{equation}
where $n_{\rm orb}$ is the number of orbits necessary for the circularization, usually ranging between 2 and 10. For
$T \sim 700$ days, we get $t_{\rm cir} \sim 1400 - 7000$ days. It is interesting to note that in the $\dot{\nu} - t$
diagram (Fig. 1), beginning from September 2005, a total period of about 1800 days can be isolated, during which the
curve shows noticeable variabilities before it finally enters the stable Slowly Spinning-Down Phase after about 2011.
This time span is roughly consistent with the $t_{\rm cir}$ value that we derived, indicating that the circularization process was in progress in this period.

\section{Conclusions and Discussion}
\label{sect:disc}

In this study, by invoking an asteroid disrupted and accreted by PSR J0738-4042, we show that the pronounced change of the spin-down rate can be
reasonably explained. Especially, the sudden initial reduction (i.e. The First Pulse in Fig. 1) of the spin-down rate comes from the initial accretion when the asteroid flies by
the central pulsar for the first time, while the subsequent drop (The Drop in Fig. 1) is due to the flying away of the bound gaseous debris. The relatively stable
spin-down rate at the last stage (the Slowly Spinning-Down Phase in Fig. 1) can be explained to be resulted from the constant accretion from the transient accretion disk formed after
the orbital circularization.

As suggested by Brook et al. (2014), we also expect the spin-down rate to return to its previous value when the material from the disrupted asteroid is exhausted. For a typical asteroid with a characteristic mass of
$7.0 \times 10^{21}~\rm g$ (with radius $\sim 60~\rm km$), the disrupted material will be entirely swallowed by PSR J0738-4042 in a timescale of
about $50(\frac{r}{60~\rm km})^{3}$ yr. However, notice that the mass range of the asteroid can extend from
about $1.0 \times 10^{10}~\rm g$ to $\sim 1.0 \times 10^{24}~\rm g$, so the exact accretion time is quite uncertain. In the future, when the spin-down rate of PSR J0738-4042 returns to its previous value, we can use the accretion timescale to constrain the mass of the asteroid.


Huang \& Geng (2014) proposed an external mechanism to explain the unprecedented anti-glitch observed in the magnetar 1E 2259+586.
In their model, the impact parameter is very small so that the solid body will collide with the neutron star before coming
to the periastron. As their collision process is very fast and violent, a glitch and an associated hard X-ray burst will be produced. While in our case, the encounter of the small body with the neutron star does not lead to a direct collision. It is a much gentler process, also greatly prolonged. The change
in the spin-down rate of PSR J0738-4042 comes from the tidal disruption and accretion of the asteroid, which is much longer and smoother compared with that happens in a direct collision.
Therefore mainly the spin-down rate of PSR J0738-4042 is affected and there is no obvious glitch induced.

In our calculations, the pericenter distance of the captured asteroid is estimated as $R_{\rm p} = 1.48 \times 10^{10}$ cm, which is
smaller than the tidal disruption radius ($R_{\rm t} = 4.9 \times 10^{10}$ cm). In this case, named as an ultraclose passage by
Rees (1988), the asteroid will be completely disrupted. As discussed by Carter \& Luminet (1982), such an asteroid
is not only elongated along the orbital direction in the process, but also will undergo compression to form a short-lived pancake aligned
in the orbital plane. They defined a factor $\beta$ ($\beta \approx R_{\rm t}/R_{\rm p}$) to derive some key parameters in
the ultraclose passage case. The maximum central temperature $\Theta_{\rm m}$ and density $\rho_{\rm m}$ can be given
by $\Theta_{\rm m} = \beta^{2}\Theta_{\ast}$ and $\rho_{\rm m} = \beta^{3}\rho_{\ast}$ respectively, where
$\Theta_{\ast}$ and $\rho_{\ast}$ are the central temperature and density of the asteroid in its unperturbed state. When
$\beta \approx 10$, the temperature and density may rise enough to detonate effectively a significant fraction of
the available thermonuclear fuel, which will affect the orbital evolution of the disrupted debris. In our case, $\beta$
is only about 3, so we believe that the orbital motion and accretion process will not be significantly modified by this effect.

\acknowledgements

\appendix
This work was supported by the National Basic Research Program of China (973 Program, Grant No. 2014CB845800), and by the National
Natural Science Foundation of China with Grant No. 11473012.

\end{document}